\documentstyle[12pt,a4wide,epsf]{article}
\def\21{$SU(2) \otimes U(1)$}
\def\321{$SU(3) \otimes SU(2) \otimes U(1)$}

\begin{document}

\newcommand{\lsim}{\mbox{\raisebox{-.9ex}{~$\stackrel{\mbox{$<$}}{\sim}$~}}}
\newcommand{\gsim}{\mbox{\raisebox{-.9ex}{~$\stackrel{\mbox{$>$}}{\sim}$~}}}

\begin{flushright}
IFIC/01--49
\end{flushright}

\bigskip

\begin{center}
{\Large\bf Modeling Quintessential Inflation}

\vspace{1.5cm}

{\large K. Dimopoulos}

{\it Physics Department, Lancaster University,\\ 
Lancaster LA1 4YB, England, U.K.}

\vspace{1cm}

{\large J. W. F. Valle }

{\it Instituto de F\'{\i}sica Corpuscular, 
C.S.I.C./Universitat de Val{\`e}ncia\\
Edificio Institutos de Paterna, Apt 22085,\\ 
E--46071 Val{\`e}ncia, Spain}

\end{center}

\vspace{1cm}

\begin{abstract} 
  
  We develop general criteria to construct unified frameworks for
  inflation and quintessence which employ a unique scalar field to
  drive both. By using such a minimal theoretical framework we avoid
  having to fine-tune couplings and mass-scales.  In particular the
  initial conditions for quintessence are already fixed at the end of
  the inflationary epoch. We provide concrete realizations of the
  method which meet all inflationary and quintessence requirements,
  such as the COBE normalization and the resulting spectral index 
  \mbox{$n \approx 0.97$}, which is in excellent agreement with the 
  latest CMB data.
\end{abstract}

\vspace{2cm}

\begin{flushleft}
{\it Key words:} inflation, quintessence, cosmological constant, CMB\\
{\it PACS:} 98.80-k,98.80.Es, 98.80.Cq, 98.80.Bp\\
\underline{\hspace{7cm}}\\
{\sl Preprint submitted to Elsevier Science}
\end{flushleft}

\pagebreak
\section{Introduction}
\label{sec:introduction}

Recent observations of remote Supernova Ia luminosity curves suggest
that the Universe is entering a second phase of accelerated expansion,
i.e. a late inflationary period~\cite{accel}.
This is also in agreement with a number of other requirements, which
suggest that the Universe at present is dominated by a dark energy
component, usually associated with the presence of a non-zero
cosmological constant \cite{dark}. Results of the latest observations of 
the Cosmic Microwave Background Radiation (CMBR) \cite{cmbobs} indicate 
that the overall energy density of the Universe corresponds to the critical
value of a spatially flat Friedman-Robertson-Walker (FRW) spacetime, 
successfully predicted by inflationary theory. In such a critical Universe 
the Cold Dark Matter (CDM) content at present cannot be larger than 35\%, 
leaving the remaining 65\% unaccounted for~\cite{Jaffe:2001tx}.
The remaining energy density is best attributed to a dark energy
component of negative pressure.  This has resurrected the embarrassing
issue of the cosmological constant $\Lambda$~\cite{Weinberg:1989cp},
previously assumed to vanish due to an unknown symmetry.  The
so--called $\Lambda$CDM scenario, which invokes a non-zero $\Lambda$ in 
order to explain the missing dark energy component, is very appealing due 
to its simplicity.  However, it is highly unpleasant from a theoretical
point of view because the required value of the cosmological constant
has to be fine tuned to the incredible level of 
\mbox{$\Lambda^2\sim 10^{-120}M_P^4$}, where $M_P$ is the Planck mass
\cite{anthropic}.

Due to this fact, the idea of a time-variable cosmological ``constant''
\cite{Peebles:1988ek} (see also \cite{early}) has become very popular because 
it can potentially account for the ``missing'' dark energy in a dynamical way.
One realization of a dynamical cosmological constant is achieved with
the use of a scalar field $Q$, assumed to dominate with its potential energy 
density the late-time history of the Universe and drive the latter into 
accelerated expansion at present. This is the so--called, QCDM scenario. 
The field $Q$ is referred to as quintessence, the fifth element in addition to 
Cold Dark Matter, Hot Dark Matter, baryons and photons \cite{quint}. 
Quintessence has inspired a large number of authors (e.g. see 
\cite{quint2}\cite{track}), who have attempted to derive it using a variety of 
theoretical frameworks such as string/M theory \cite{string-q}, scalar-tensor 
gravity \cite{nmc}, brane-worlds and large extra dimensions \cite{led}, 
supersymmetry and supergravity \cite{susy-q}\cite{exp-q}. However, 
it has been soon realized that quintessence has to be fine-tuned in a way 
similar to $\Lambda$ itself \cite{Kolda:1999wq}. Indeed, while at present 
\mbox{$Q\sim M_P$}, its mass should be \mbox{$m_Q\sim 10^{-33}$eV}!
It is clearly not an easy task to achieve such extremely small mass
from a potential $V(Q)$ without fine-tuning mass-scales or couplings.
Moreover, the introduction of yet again another unobserved scalar
field, whose origin is unaccounted for, seems unappealing.  Finally, a
rolling scalar field introduces another fine-tuning problem, namely
that of its initial conditions.

A compelling way to overcome the difficulties and disadvantages of the 
QCDM scenario is to link it with the rather successful inflationary theory.
Linking inflation and quintessence is rather natural because both theories 
are based on the same idea, namely that the Universe undergoes a phase of 
accelerated expansion when dominated by the energy density of a scalar field, 
which slowly rolls down its almost flat potential. The successes of 
inflationary theory are many. Indeed, inflation provides so far the only 
solution for the horizon and flatness problems compatible with the 
cosmological principle~\cite{Albrecht:1982wi}, while also giving a natural 
explanation for the formation of Large Scale Structure (LSS) (i.e. the 
distribution of galactic clusters and superclusters) as well as the amplitude 
and Doppler peaks of the spectrum of the anisotropies of the 
CMBR~\cite{deBernardis:2000gy}. Finally, the cosmological parameters inferred 
from present observations are totally consistent with inflationary initial 
conditions.

Quintessential inflation is achieved by identifying 
$Q$ with the inflaton field $\phi$. Thus, in such a model, the form of the 
scalar potential is such that results in two phases of accelerated expansion,
one at early and the other at late times.
There are several possibilities as for the theoretical origin of this
scalar field.  For example, supergravity theories provide a natural 
setting for inflationary models~\cite{Lyth:1999xn} as well as quintessence
models (e.g. Copelend {\it et al.} or Brax and Martin in \cite{susy-q}). 
Hence, it is reasonable to adopt such a framework also to model 
quintessential inflation as suggested here. 

The attempt to use inflation to account for the present-day vacuum
energy~\cite{Frewin:1993aj} has resulted in relatively few models of
quintessential inflation to date \cite{q-inf-1} - \cite{q-inf-branes}. 
The rarity of quintessential inflationary models in the literature reflects 
the fact that the formulation of the required scalar potential is very
difficult. Indeed, a successful model of quintessential inflation is
subject not only to the requirements of inflation and quintessence,
but also to a number of additional considerations.
For example, the minimum of the potential (taken to 
be zero, otherwise there is no advantage over $\Lambda$CDM) must not have 
been reached yet by the rolling scalar field. This requirement is typically 
satisfied by potentials, which have their minimum displaced at infinity, 
\mbox{$V(\phi\!\to\!\infty)\to 0$}, a feature referred to as ``quintessential 
tail''. Thus, quintessential inflation is a non-oscillatory inflationary 
model \cite{Felder:1999pv}.
Another requirement is that of a ``sterile'' inflaton, i.e. the scalar
$\phi$ should not couple directly to any fields of the Standard Model
(SM).  This is necessary in order to avoid the decay of $\phi$ 
into a thermal bath of SM particles at the end of inflation, since 
we need $\phi$ to survive until today, as required in order to drive the 
present late phase of accelerated expansion. An additional advantage of 
considering a ``sterile'' inflaton is that one avoids the fine-tuning
of the couplings between the inflaton and the SM particles, typical for
usual inflationary models, in which these couplings have to be kept small 
in order to preserve the flatness of the inflationary potential. Moreover,
a sterile $\phi$ avoids the violation of the equivalence principle at present,
due to the fact that the ultra-light quintessence field would otherwise
correspond to a long-range force. 
Natural candidates for such sterile inflaton are moduli fields, hidden 
sector and mirror fields, the radion and so on. In the absence of any
couplings with SM fields $\phi$ continues its roll-down after the end of
inflation, whereas the Universe reheats via the gravitational
production of particles at the end of the inflationary
period~\cite{Ford:1987sy}\cite{Joyce:1998fc}. Because gravitational reheating 
is inefficient the Universe remains $\phi$-dominated after the end of 
inflation, this time by the kinetic energy of the scalar field
\cite{Joyce:1998fc}\cite{Spokoiny:1993kt}.
This period of kinetic energy domination, however, is terminated quickly
and the Universe enters the radiation domination period of the Standard 
Hot Big Bang (SHBB).

In the models of \cite{q-inf-1} and \cite{q-inf-2} the plethora of constraints 
and requirements which are to be satisfied by quintessential inflation is 
managed through the use of a multi-branch scalar potential, that is a 
potential which changes its form while the field moves from the inflationary 
to the quintessential phase of its evolution. This change is either fixed 
``by hand'' (such as in the toy-model of \cite{q-inf-1}) or it is the outcome 
of a phase transition, which is arranged through some interaction of the 
inflaton with some other scalar fields (as in the case of \cite{q-inf-2}). 
Clearly this requires the introduction of additional scalar fields. Moreover, 
models which involve a phase transition also need many model parameters, 
couplings and mass scales that have to be tunned correctly to achieve the 
desired results. Thus, in such models it is difficult to dispense with the 
fine-tuning problems of quintessence. Still, a remarkable achievement of the 
phase-transition models of \cite{q-inf-2} is that they manage to formulate the 
scalar potential for quintessential inflation in the context of supersymmetry. 
Another interesting approach is the one of \cite{Kaganovich:2001fc}, where a 
large variety of possible choices for the quintessential inflationary 
potential is generated through a theory of non-minimally coupled gravity. 
However, the author of \cite{Kaganovich:2001fc} is primarily interested in 
demonstrating the ability of this framework to allow for the construction of 
many types of successful potentials without actually attempting to construct 
models with minimal paramater content, although the latter is plausible in 
this theory. Finally, although rather interesting, recent attempts to model 
quintessential inflation using brane-world physics \cite{q-inf-branes} cannot 
avoid fine-tunning problems due to the large number of undetermined free 
parameters inherent to the brane-environment.

In this paper we develop general criteria and provide concrete
realizations of unified frameworks for inflation and quintessence
driven by a single scalar field $\phi$. We adopt an alternative approach to 
the problem and attempt to formulate quintessential inflation in terms of a 
single-branch scalar potential, which communicates information from the 
inflationary to the quintessential phase of the scalar field's evolution. 
By using such minimal theoretical framework, we avoid introducing additional
scalars and minimize the fine-tuning of couplings and mass-scales.
For example, the initial conditions for quintessence are already fixed
at the end of the inflationary epoch. 

Our paper is organized as
follows. In Sec.~\ref{sec:infl-quint} we discuss, in a
model--independent way, the main requirements of quintessential inflation, 
i.e. those that follow from inflation, such as
slow--roll, density perturbations and correct amplitude and spectrum
of CMBR anisotropies, sufficient reheating for nucleosynthesis,
as well as those that follow from quintessence, such as accelerated
expansion at present. We also introduce the concepts of frozen and
attractor quintessence.  In Sec.~\ref{sec:how-model-infl} we present
some general criteria required in order to model quintessential
inflation, which suggest a preferred choice of the form of the quintessential 
tail of the scalar potential. Then, in Sec.~\ref{sec:examples} we present a 
concrete realization in terms of a specific form of the potential and
demonstrate how it meets all the necessary requirements outlined
previously in Secs.~\ref{sec:infl-quint} and \ref{sec:how-model-infl}.
Finally, in Sec.~\ref{sec:disc-concl} we draw our conclusions and provide
additional discussion on the method and examples presented. Throughout our 
paper we use units such that \mbox{$c=\hbar=1$} in which Newton's 
gravitational constant is \mbox{$G=M_P^{-2}$}, where 
\mbox{$M_P=1.22\times 10^{19}$GeV}.

\section{Quintessential Inflation}
\label{sec:infl-quint} 

\subsection{The dynamics of the Universe}
\label{sec:dynamics-universe}

The equations of motion for a spatially flat FRW Universe consist of
Friedman equation, the energy-momentum conservation condition and
Raychadhuri equation, as follows
\begin{eqnarray}
H^2 & = & \frac{\rho}{3m_P^2}  \label{H}\\
 & & \nonumber\\
d(a^3\rho) & = & -p\,d(a^3) \label{emc}\\
 & & \nonumber\\
\frac{\ddot{a}}{a} & = & -\frac{\rho+3p}{6m_P^2} \label{ray}
\end{eqnarray}
where \mbox{$m_P\equiv M_P/\sqrt{8\pi}$} is the reduced Planck mass, $\rho$ 
and $p$ are the total energy density and pressure of the Universe,
\mbox{$H\equiv\dot{a}/a$} is the Hubble parameter with $a(t)$ being the
scale factor of the Universe and the dot denotes derivative with respect to 
the cosmic time $t$.  The content of the Universe is modeled as a number
of perfect-fluid components with equations of state
\mbox{$p_i=w_i\rho_i$}, so that \mbox{$\rho=\sum_i\rho_i$} and 
\mbox{$p=\sum_ip_i$}.

For each of these components (\ref{emc}) suggests, 
\begin{equation}
\rho_i\propto a^{-3(1+w_i)}
\label{r-a}
\end{equation}

If, during a particular period of the Universe evolution, one of the
density components, say $\rho_D$, dominates we can identify 
\mbox{$\rho\simeq\rho_D$} and \mbox{$p\simeq p_D\equiv w_D\rho_D$} so that, 
using (\ref{H}) and (\ref{r-a}) we find,,

\begin{eqnarray}
H=\frac{2\,t^{-1}}{3(1+w_D)} & 
\hspace{1cm}
a\propto t^{^{\;\mbox{\footnotesize $\frac{2}{3(1+w_{\!D})}$}}}
\hspace{1cm}
 & \rho=\frac{4}{3(1+w_D)^2}\Big(\frac{m_P}{t}\Big)^2
\label{wD}
\end{eqnarray}
where \mbox{$w_D\neq -1$}. In the case of a cosmological constant
domination \mbox{$\rho_\Lambda=m_P^2\Lambda=$ const.} and, from (\ref{r-a}) 
we have
\mbox{$w_\Lambda=-1$} so that (\ref{H}) gives \mbox{$H^2=\frac{1}{3}\Lambda$} 
and \mbox{$a\propto\exp(Ht)$}, i.e. the Universe undergoes pure de-Sitter
inflation.

We will assume that \mbox{$\Lambda=0$} and the Universe is filled with a 
background density $\rho_B$ 
comprised by pressureless matter $\rho_m$ (including baryons and CDM) with 
\mbox{$w_m=0$} and radiation $\rho_\gamma$ (including relativistic matter) 
with \mbox{$w_\gamma=1/3$} so that,

\begin{equation}
p_B=w_B\rho_B
\end{equation}
where $w_B=1/3$ \{0\} for the radiation \{matter\} era. We also add to the
background a homogeneous scalar field \mbox{$\phi=\phi(t)$}, which can also
be treated like a perfect fluid\footnote{We will not be concerned here with 
inhomogeneities of this scalar field due to fluctuations, because their 
magnitude is much smaller than the homogeneous mode.} with,

\begin{equation}
\label{phi}
\begin{array}{lll}
\rho_\phi\equiv\frac{1}{2}\dot{\phi}^2+V(\phi) & 
\hspace{0.5cm}\mbox{and}\hspace{0.5cm} &
p_\phi\equiv\frac{1}{2}\dot{\phi}^2-V(\phi)
\end{array}
\end{equation}
where \mbox{$V=V(\phi)$} is the scalar potential of $\phi$. From the latter
it follows that

\begin{equation}
w_\phi=\frac{\dot{\phi}^2-2V(\phi)}{\dot{\phi}^2+2V(\phi)}
\label{wphi}
\end{equation}

From the above, (\ref{ray}) suggests that the Universe may undergo
accelerated expansion only if $\rho_B<\rho_\phi$ and \mbox{$w_\phi<-\frac{1}{3}$}.
In order to follow the Universe dynamics during the $\phi$-dominated
periods we also need the scalar field equation of motion,

\begin{equation}
\ddot{\phi}+3H\dot{\phi}+V'=0 
\label{field}
\end{equation}
where the prime denotes derivative with respect to $\phi$. Finally, the
temperature of the Universe is given by,
\begin{equation}
\rho_\gamma=\frac{\pi^2}{30}g_* T^4
\label{T}
\end{equation}
where $g_*$ is the number of relativistic degrees of freedom, which for the 
standard model in the early Universe, is \mbox{$g_*=106.75$}.

\subsection{Inflation}
\label{sec:inflation}

Inflation occurs when the Universe is $\phi$-dominated and the form of
$V(\phi)$ is such that the so--called slow--roll parameters
\begin{equation}
\varepsilon\equiv\frac{m_P}{\sqrt{6}}\frac{V'}{V}\hspace{1.5cm}
\eta\equiv\frac{m_P^2}{3}\frac{V''}{V}
\label{eps}
\end{equation}
satisfy the slow--roll conditions: $|\varepsilon|,|\eta|<1$.  In view of these
conditions (\ref{H}) and (\ref{field}) become,

\begin{eqnarray}
H^2 & \simeq & V/3 m_P^2 \label{Hinf}\\
 & & \nonumber\\
3H\dot{\phi} & \simeq & -V' \label{infeqm}
\end{eqnarray}

In general, during inflation it is useful to express all the relevant 
quantities as functions of the number of remaining e-foldings until the end of 
inflation, which is defined as,

\begin{eqnarray}
N\equiv\int^{a_N}_{a_{\rm end}}\frac{da}{a}\simeq
\int_{\phi_{\rm end}}^{\phi_N}\frac{V}{V'}\frac{d\phi}{m_P^2}=
\frac{1}{\sqrt{6}}\int_{\phi_{\rm end}}^{\phi_N}\frac{d\phi}{\varepsilon\,m_P}
\label{N}
\end{eqnarray}

Values of $N$ that are of particular interest have to do with length
scales that exit the Horizon $N$ e-foldings before the end of
inflation and re-enter during the Standard Hot Big Bang (SHBB) at
important moments of the evolution of the Universe. Suppose one is
interested on the length scale that re-enters the Horizon at time
$t_N$. The corresponding $N$ is then,

\begin{equation}
N\simeq\ln\Big(\frac{T_N}{H_N}\Big)+
\ln\Big(\frac{H_{\rm end}}{T_{\rm reh}}\Big)
\label{NTH}
\end{equation}
where \mbox{$T_N\equiv T(t_N)$} and \mbox{$H_N=H(t_N)$} are 
the temperature and the Hubble parameter of the Universe respectively
at the time of re-entry, 
$T_{\rm reh}$ is the reheating temperature and
\mbox{$H_{\rm end}\equiv H(t_{\rm end})$} is the 
Hubble parameter at the end of inflation
\mbox{$t_{\rm end}\sim m_P/T_{\rm reh}^2$}.

\subsubsection{Density perturbation  requirements}
\label{sec:infl-requ}

One of the big achievements of the inflationary universe idea is that
it provides a natural origin for almost scale invariant cosmological
density perturbations~\cite{Lyth:1999xn}.  The amplitude of the
density perturbations is,
\begin{equation}
\frac{\delta\rho}{\rho}\;\simeq\;
\left.\frac{H^2}{\pi\dot{\phi}}\,\right|_{\rm exit}
\simeq -\frac{1}{\sqrt{3}\pi}\frac{V^{3/2}}{m^3_PV'}
\label{dr/r}
\end{equation}
where ``exit'' denotes that the ratio $H^2/\pi\dot{\phi}$ should be
evaluated when the scale of interest exits the horizon during
inflation.  The scale of interest is the one which reenters the
horizon at the moment of decoupling between matter and radiation
\mbox{$t_{\rm dec}\sim 10^{12}$sec} (when the CMBR is emitted) because
the corresponding overdensity contrast is constrained by COBE
observations. The COBE constraint reeds,
\begin{equation}
\left.\frac{\Delta T}{T}\right|_{\rm dec}\simeq
\frac{\delta\rho}{\rho}(N_{\rm dec})\simeq 2\times 10^{-5}
\label{cobe}
\end{equation}
where $N_{\rm dec}$ is the remaining number of e-foldings of inflation when 
the scale in question exits the horizon.

The spectral index of the overdensity spectrum is defined as, 
\begin{equation}
n\equiv 1+\frac{d\ln P_k}{d\ln k}
\end{equation}
where \mbox{$k^{-1}\sim He^N$} is the relevant scale at the end of
inflation and, \mbox{$(P_k)^{1/2}\sim\frac{\delta\rho}{\rho}(k)$}. After some
algebra one finds,
\begin{equation}
n-1
\simeq 6(\eta-3\varepsilon^2)
\label{n}
\end{equation}

Large Scale Structure (LSS) and CMBR observations~\cite{deBernardis:2000gy}
suggest that,
\begin{equation} 
|n(N_c)-1|\leq 0.1
\end{equation}
where \mbox{$N_c=N_{\rm dec}, N_{\rm eq}$} with $N_{\rm eq}$ being the
remaining number of e-foldings when the scale which reenters the
horizon at $t_{\rm eq}$ exits the horizon during inflation, where 
\mbox{$t_{\rm eq}\sim 10^{11}$sec} is the time of equal matter
and radiation densities (when the main features of LSS are determined).
This requirement is achieved if both the slow--roll conditions are strongly
satisfied when \mbox{$N=N_c$} (or if \ 
\mbox{$3\varepsilon^2(N_c)\approx\eta(N_c)$} such
as in \cite{Dimopoulos:2001md}).

\subsubsection{Horizon and flatness  requirements}
\label{sec:horizon-flatness}

The horizon problem is solved if the present horizon scale did exit
the horizon during the inflationary period. The e-foldings $N_H$ since
when this happened are found with the use of (\ref{NTH}),
\begin{equation}
N_H\simeq\ln(T_{\rm\sc cmb}t_0)+\ln(\frac{H_{\rm end}}{T_{\rm reh}})
\label{NH}
\end{equation}
where $T_{\rm\sc cmb}$ is the CMBR temperature at present and $t_0$ is
the SHBB age of the Universe. Solving the flatness problem requires
the total number of e-foldings to be large enough to make
\mbox{$\Omega\leadsto 1$}, where \mbox{$\Omega\equiv\rho/\rho_c$}, with 
$\rho_c$ being the critical energy density for a spatially flat Universe. This
requirement is quantified if we assume an initial $\Omega-1$ value of order
unity, in which case we find that the necessary number of inflationary
e-foldings is,

\begin{equation}
N_F\simeq\frac{1}{2}\Big[\ln\Big(\frac{t_{\rm eq}}{t_{\rm end}}\Big)+
\frac{2}{3}\ln\Big(\frac{t_{\rm dec}}{t_{\rm eq}}\Big)
-\ln|\Omega_{\rm dec}-1|\Big]
\label{NF}
\end{equation}
where \mbox{$\Omega_{\rm dec}\equiv\Omega(t_{\rm dec})$}. 
The above can be evaluated using the results of the CMBR 
observations~\cite{cmbobs} which suggest that,

\begin{equation}
|\Omega_{\rm dec}-1|\leq 0.1
\end{equation}

Thus, the Horizon and flatness problems are solved if,

\begin{equation}
N_{\rm tot}\geq\mbox{max}\{N_H,N_F\}
\end{equation}

Typically the density perturbation requirements are used to evaluate the 
parameters of an inflationary model whereas the horizon and flatness 
requirements are constraints on the inflationary initial conditions.

\subsection{After the end of inflation}
\label{sec:after-end-inflation}

\subsubsection{Reheating}

In quintessential inflation the inflaton does not decay into a thermal
bath of Standard Model particles so that it may survive until today
and act as quintessence. To that end $\phi$ does not oscillate around 
its minimum at the end of inflation but continues to roll-down its potential,
violating however, the slow roll conditions, so that inflation is terminated 
(see Fig. 1). The necessary particle (and entropy) production occurs by
gravitational means~\cite{Ford:1987sy}. Since the gravitationally
generated fields have an approximately scale invariant spectrum with
amplitude given by the Hawking temperature, we have,
\begin{equation}
T_{\rm reh}\simeq 
\alpha\left(\frac{H_{\rm end}}{2\pi}\right)
\label{Trehg}
\end{equation}
where \mbox{$\alpha\sim 0.01$} is an efficiency factor \cite{Joyce:1998fc}. 
In view of this we have \mbox{$(H_{\rm end}/T_{\rm reh})=2\pi/\alpha$}. 
Thus, from (\ref{NTH}) it is evident that, in the case of gravitational
reheating, the number of e-foldings before the end of inflation that
corresponds to a length scale that re-enters the Horizon at a
particular time $t_N$ is independent of the inflationary energy scale
and, indeed, of any of the model parameters.  The only dependence is
on the reheating efficiency factor $\alpha$. Thus, in particular, for the
scale of the Horizon today and the scales that re-enter the Horizon at
$t_{\rm dec}$ and $t_{\rm eq}$ respectively we find,

\begin{eqnarray}
N_H & \simeq & 69.15-\ln\alpha\nonumber\\
N_{\rm dec} & \simeq & 66.94-\ln\alpha\label{Ns}\\
N_{\rm eq} & \simeq & 64.27-\ln\alpha\nonumber
\end{eqnarray}

\begin{figure}
\begin{center}
\leavevmode
\hbox{%
\epsfxsize=4.0in
\epsffile{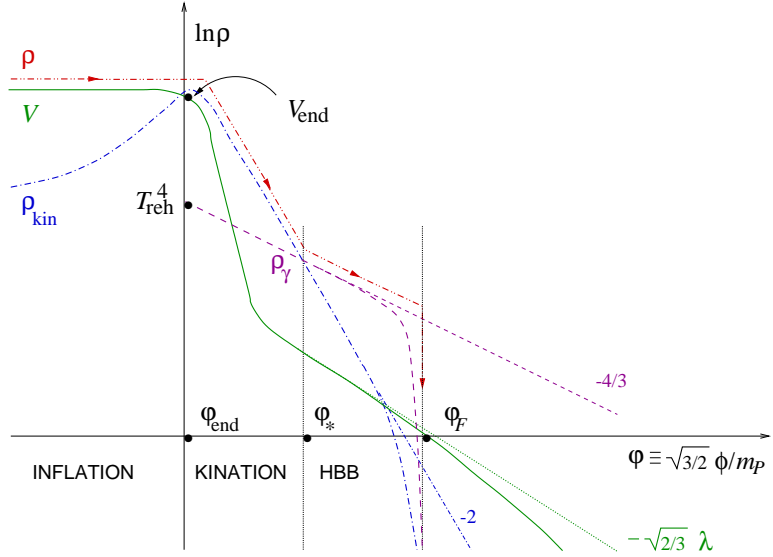}}
\end{center}

\caption{\footnotesize 
Schematic view of the scalar potential in Quintessential Inflation 
(solid line). The potential features two flat regions, the inflationary 
plateau and the quintessential tail. The global minimum is at $+\infty$ so 
that \mbox{$V(+\infty)=0$}. With such potential inflation is terminated by a 
drastic reduction of the potential energy density, leading to a rapid 
roll-down of the scalar field from the inflationary plateau towards the 
quintessential tail. During inflation \mbox{$\rho_{\rm kin}=\varepsilon^2V$} 
so that, at the end of inflation (\mbox{$\varepsilon\simeq 1$}), we have 
\mbox{$V_{\rm end}\simeq\rho_{\rm kin}(\phi_{\rm end})$}.
Due to this fact, after the end of inflation, the kinetic energy density 
of the scalar field $\rho_{\rm kin}$ (dash-dot line) dominates for a brief 
period the energy density of the Universe. During this kination phase the 
background energy density \mbox{$\rho_B\simeq\rho_\gamma$} (dashed line) 
reduces less rapidly than $\rho_{\rm kin}$. As a result, even though at the 
end of inflation 
\mbox{$\rho_\gamma(\phi_{\rm end})\sim T_{\rm reh}^4\ll V_{\rm end}$},
beyond a certain value $\phi_*$ of the scalar field, the Universe becomes 
radiation dominated and the Hot Big Bang (HBB) begins. After that the kinetic 
energy of the scalar filed reduces rapidly to zero (i.e. 
\mbox{$\rho_{\rm kin}\rightarrow 0$}) and the field (asymptotically) freezes 
to a value $\phi_F$, while the overall density of the Universe $\rho$ 
(dash-dot-dot line) continues to decrease due to Hubble expansion. In 
the above figure we have plotted the energy densities of the various 
components of the Universe, as well as the overall energy density of 
the Universe, with respect to the dimensionless quantity 
\mbox{$\varphi\equiv\sqrt{3/2}\,\phi/m_P$}. Note that in a logarithmic plot, 
it is easier to depict the difference of the scaling of the various 
components of the Universe density. For example, during kination, it can be 
shown that \mbox{$t\propto e^\varphi$} so that 
\mbox{$\rho_{\rm kin}\propto e^{-2\varphi}$} and 
\mbox{$\rho_\gamma\propto e^{-\frac{4}{3}\varphi}$}. Assuming a 
quasi-exponential quintessential tail of the form given by (\ref{tail})
we have \mbox{$V\propto e^{-\sqrt{2/3}\,\lambda\varphi}/\varphi^k$}. Hence, 
the potential energy density in the above plot falls logarithmicaly away
from the pure-exponential case (dotted line). In the HBB 
era it can be shown that \mbox{$t\propto (\varphi_F-\varphi)^{-2}$} so that
\mbox{$\rho_{\rm kin}\propto (\varphi_F-\varphi)^6$} and 
\mbox{$\rho_\gamma\propto (\varphi_F-\varphi)^4$}. Consequently, in the 
above plot, \mbox{$\rho_{\rm kin},\rho_\gamma\rightarrow 0$} while 
\mbox{$\phi\rightarrow \phi_F$}.}
\label{q1}
\end{figure}

\subsubsection{Kination}
\label{sec:kination}

After the end of inflation the Universe is dominated, for some time, by the 
kinetic energy of the scalar field. Such a period is refereed to as kination 
or deflation~\cite{Joyce:1998fc}\cite{Spokoiny:1993kt}. 
During this epoch \mbox{$w_D=w_\phi=1$} (c.f. (\ref{wphi})~) and, from 
(\ref{r-a}), 
\mbox{$\rho_{\rm kin}\equiv\frac{1}{2}\dot{\phi}^2\propto a^{-6}$}, 
i.e. the kinetic energy of $\phi$ decreases faster than $\rho_B$.
Thus, in due time $\rho_B$ will catch up with $\rho_{\rm kin}$ and kination
will be terminated (see Fig. \ref{q1}).

From (\ref{wD}) we have \mbox{$a\propto t^{1/3}$} and \mbox{$H=1/3t$}.  In 
view of this and using, (\ref{infeqm}) and (\ref{Trehg}) we find that 
kination ends at,
\begin{equation}
t_*=\big(\frac{30}{\pi^2g_*}\big)^{3/2} 
t_{\rm end}\,\frac{V_{\rm end}^{3/2}}{T_{\rm reh}^6}=
\frac{1}{9}\Big(\frac{6\sqrt{\pi}}{\alpha}\Big)^6
\Big(\frac{10}{g_*}\Big)^{3/2}
\frac{m_P^7}{V_{\rm end}^2}
\label{t*}
\end{equation}
where, in view of (\ref{wD}), \mbox{$t_{\rm end}= m_P/\sqrt{3V_{\rm end}}$} 
is the time when inflation ends and \mbox{$V_{\rm end}\equiv V(t_{\rm end})$}.

The corresponding temperature is obtained with the use of (\ref{wD}) and 
(\ref{T}),

\begin{equation}
T_*=\pi\sqrt{\frac{g_*}{30}}\,\frac{T_{\rm reh}^3}{V_{\rm end}^{1/2}}=
\frac{\alpha^3}{72\pi^2}\sqrt{\frac{g_*}{30}}\,\frac{V_{\rm end}}{m_P^3}
\label{T*}
\end{equation}

Note that we require $t_*$ to be before Big Bang Nucleosynthesis
(BBN). Thus,
\begin{equation}
T_{\rm\sc bbn}<T_*
\label{bbn}
\end{equation}
where \mbox{$T_{\rm\sc bbn}\simeq 0.5$ MeV} is the temperature at the onset
of BBN. Note that, in view of (\ref{T*}), the above constraint provides a 
lower bound on $V_{\rm end}$.

During kination the scalar field equation of motion (\ref{field}) is written 
as,
\begin{equation}
\ddot{\phi}+(\dot{\phi}/t)\simeq 0
\label{kinfield}
\end{equation}
Solving we find that the scalar field rolls down until the value,
\begin{equation}
\phi_*\equiv\phi(t_*)\simeq\phi_{\rm end}+\sqrt{6}\,\left[
2\,\ln\Big(\frac{m_P}{V^{1/4}_{\rm end}}\Big)+
\ln\Big(\frac{12\pi}{\alpha^2}
\mbox{\Large $\sqrt{\frac{30}{g_*}}$}
\,\Big)\right]m_P
\label{f*}
\end{equation}
where \mbox{$\phi_{\rm end}\equiv\phi(t_{\rm end})$} is the value of $\phi$ at 
which inflation is terminated.

\subsubsection{Hot Big Bang}
\label{sec:hot-big-bang}

At the onset of the radiation era $\rho_{\rm kin}$ decreases rapidly as,
\mbox{$\rho_{\rm kin}\propto t^{-3}$} since \mbox{$a\propto t^{1/2}$}. Thus $\dot{\phi}\to
0$ in a short time and the field freezes at a value $\phi_F$.  Indeed,
during the radiation era, \mbox{$H=1/2t$} and (\ref{kinfield})
suggests,

\begin{equation}
\phi(t)=\phi_*+2\dot{\phi_*}t_*\Big(1-\sqrt{\frac{t_*}{t}}\,\Big)
\label{frd}
\end{equation}
where \mbox{$\dot{\phi}_*\equiv\dot{\phi}(t_*)$} is the value of $\dot{\phi}$ at the
onset of radiation domination $t_*$.  From the above we see that for
\mbox{$t>t_*$} the value of $\phi$ approaches a freezing
point~\footnote{Just before freezing $\rho_\phi$
  becomes again $V$-~dominated but by that time $\phi$ has rolled so
  much down-hill that $V$ is very small. Thus  the
  value of $\phi_F$ should not be substantially affected.},
\mbox{$\phi_F\simeq\phi_*+2\dot{\phi_*}t_*$},

\mbox{$\phi_F\simeq\phi_*+2\dot{\phi_*}t_*$}.

For gravitational reheating, (\ref{frd}) gives,
\begin{equation}
\phi_F\simeq\phi_{\rm end}+2\sqrt{\frac{2}{3}}
\left[1+\frac{3}{2}\ln\Big(\frac{12\pi}{\alpha^2}
\mbox{\Large $\sqrt{\frac{30}{g_*}}$}
\,\Big)+3\ln\Big(\frac{m_P}{V_{\rm end}^{1/4}}\Big)\right]\,m_P
\label{ffgr}
\end{equation}
where we have used (\ref{f*}) and that, during kination,\footnote{Without  
loss of generality we assume that \mbox{$\dot{\phi}\geq 0$}.}
\mbox{$\dot{\phi}=\sqrt{\frac{2}{3}}\,m_P/t$}.

\subsection{Quintessence}
\label{sec:quintessence}

\subsubsection{Frozen and attractor quintessence}
\label{sec:froz-attr-quint}

After the freezing of the scalar field its energy density is again
$V$-dominated. Thus, its evolution is described by~(\ref{infeqm}), in which, 
however, \mbox{$H^{-1}=\frac{3}{2}(1+w_B)t$} as suggested by (\ref{wD}),
because the Universe is now dominated by $\rho_B$. The solution of the
equation of motion is of the form,

\begin{equation}
f(\phi\mbox{\small $(t)$})-f(\phi_F)=F(t)
\end{equation}
where
\begin{eqnarray}
F(t)\equiv\frac{1}{4}(1+w_B)t^2 & \hspace{1cm}\mbox{and}\hspace{1cm} &
f(\phi)\equiv-\int'\frac{d\phi}{V'(\phi)}
\label{f}
\end{eqnarray}
where the prime on the integral means that one should not consider constants 
of integration.

\begin{figure}
\begin{center}
$\begin{array}{cc}
\hspace{-1.5cm}
\leavevmode
\hbox{
\epsfxsize=3.4in
\epsffile{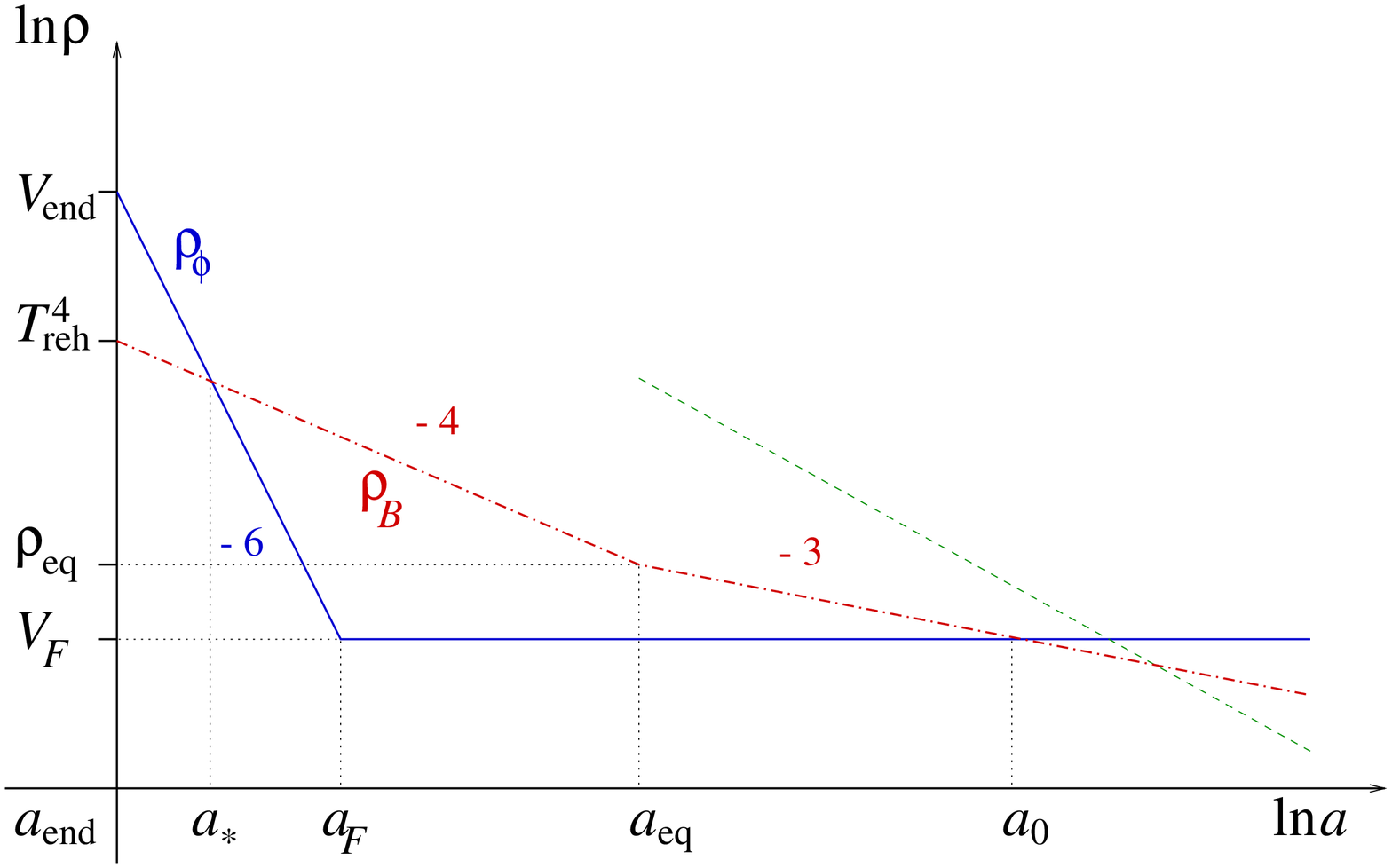}}^{\hspace{-0.6cm}(a)} &
\leavevmode
\hbox{
\epsfxsize=3.4in
\epsffile{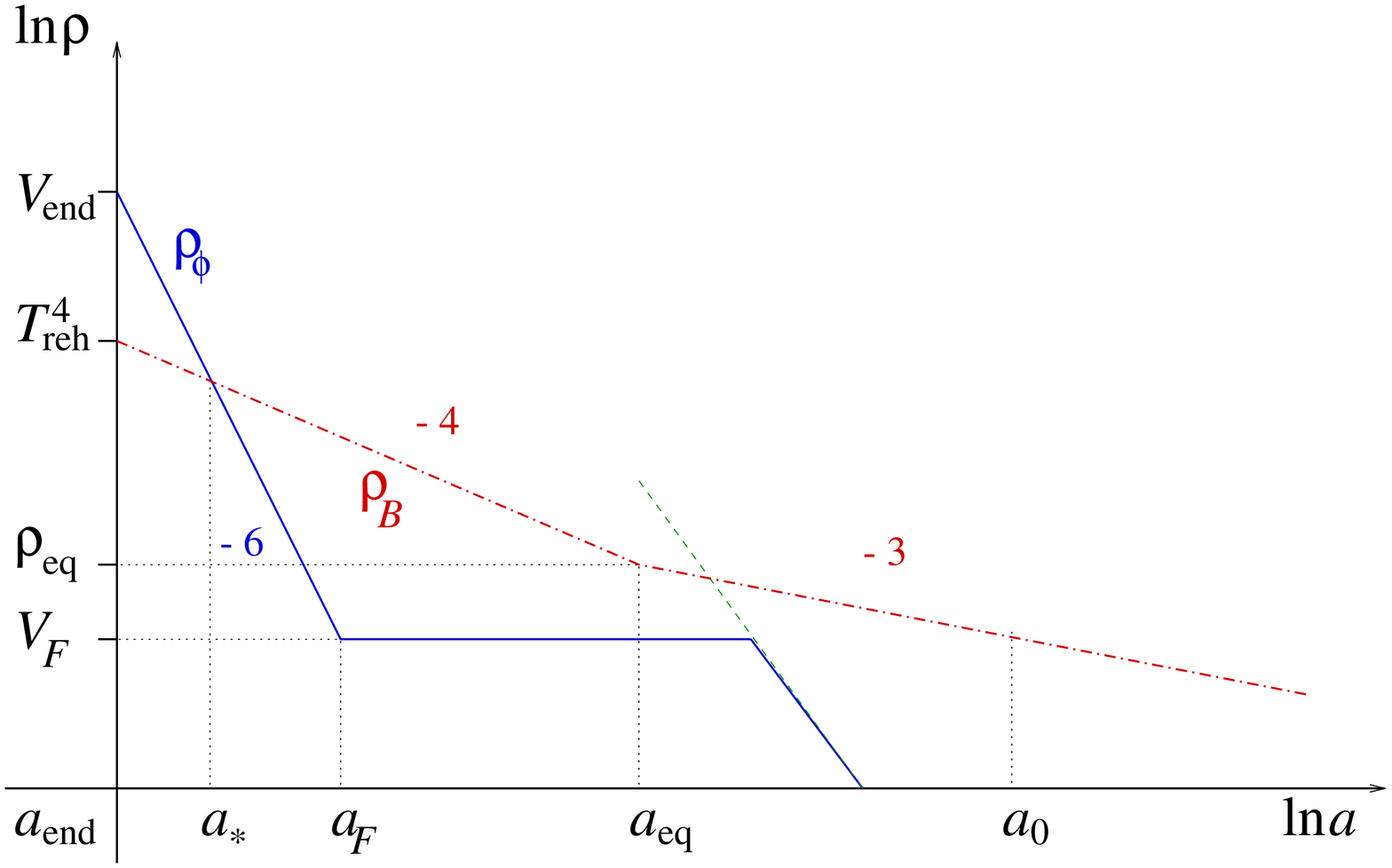}}^{\hspace{-0.6cm}(b)}
\\
 & \\
\hspace{-1.5cm}
\leavevmode
\hbox{
\epsfxsize=3.4in
\epsffile{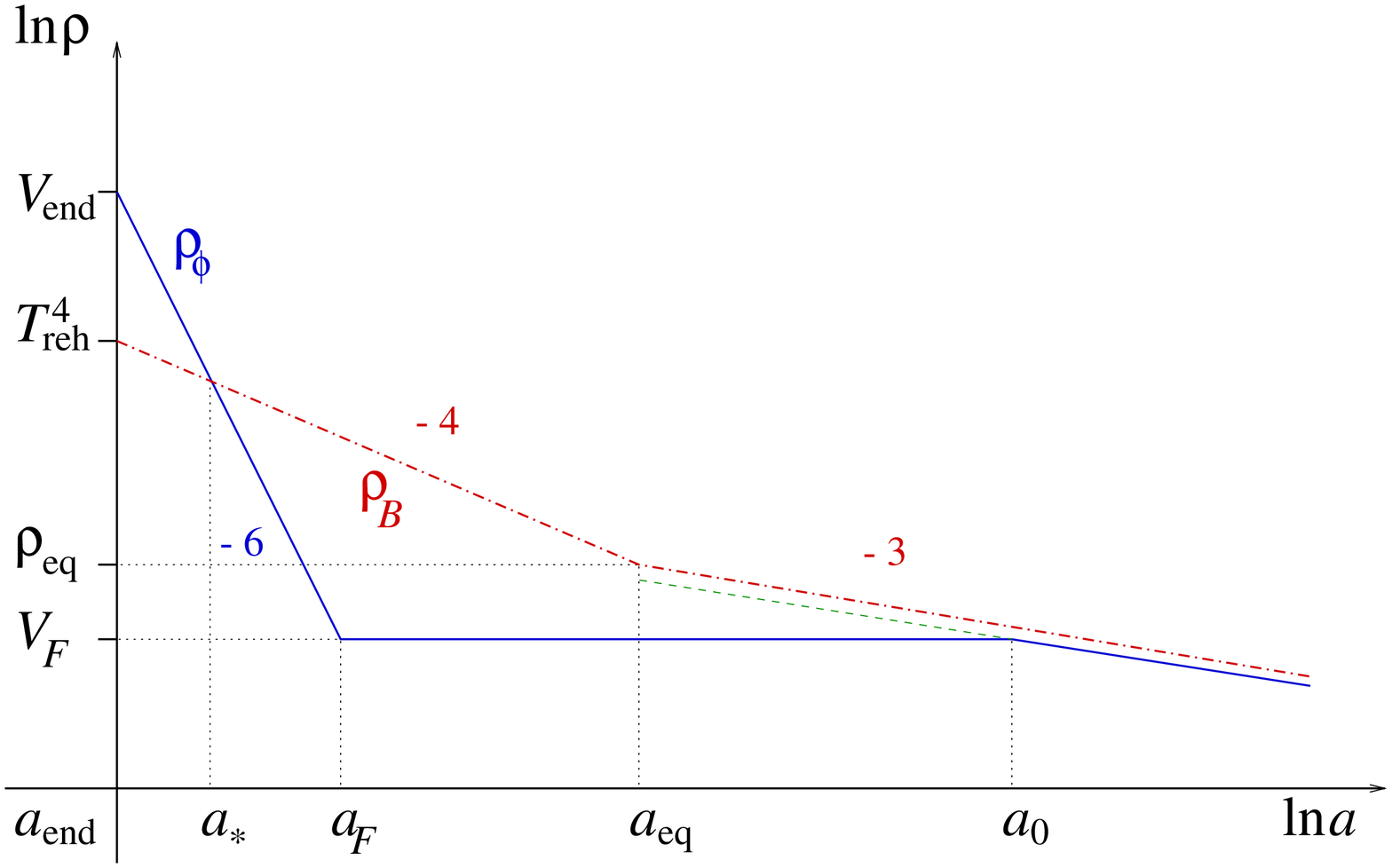}}^{\hspace{-0.6cm}(c)} &
\leavevmode
\hbox{
\epsfxsize=3.4in
\epsffile{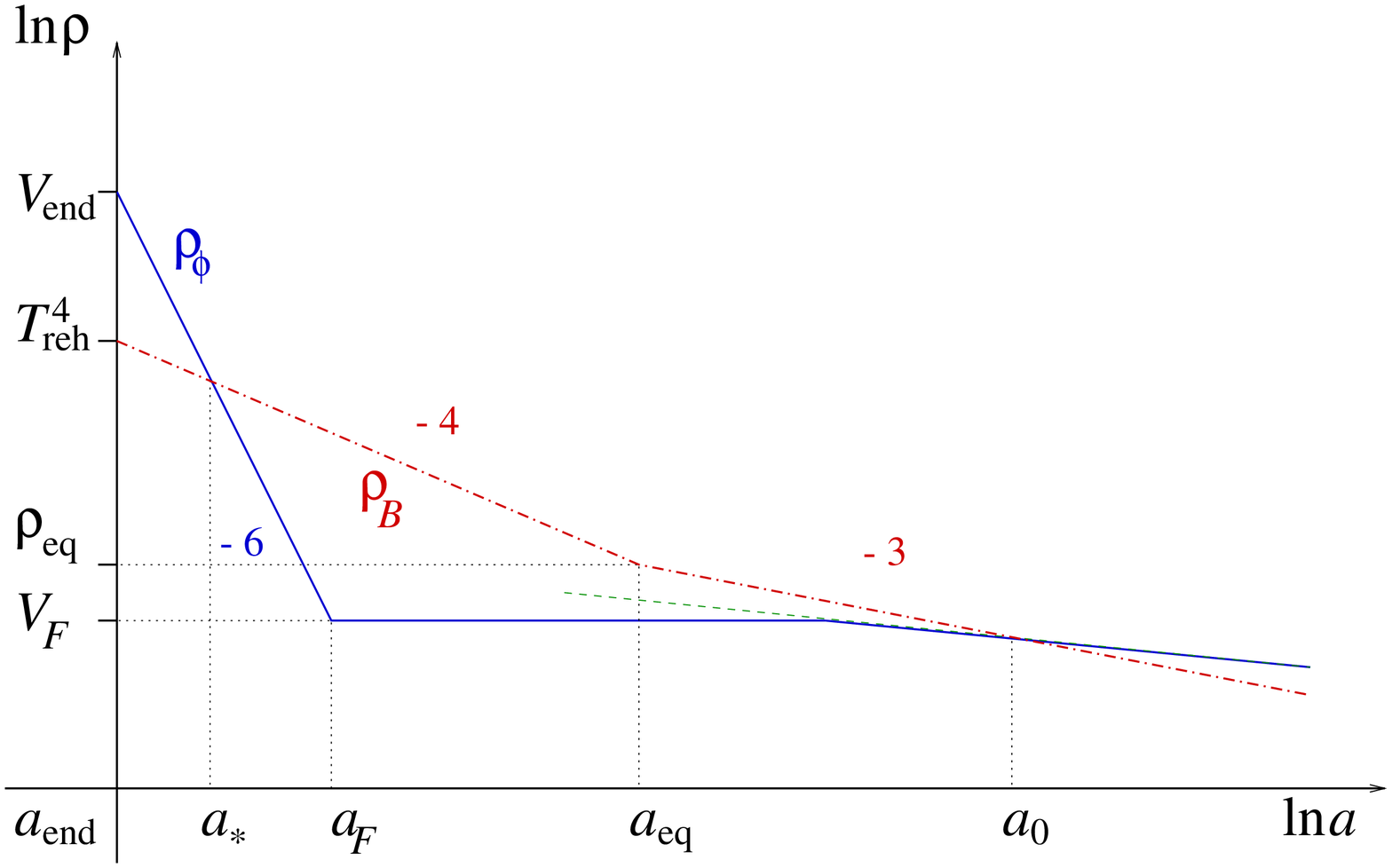}}^{\hspace{-0.6cm}(d)}
\end{array}$
\end{center}

\caption{\footnotesize
Diagrams depicting the behaviour of the energy density of the scalar field 
$\rho_\phi$ (solid line) and the background energy density $\rho_B$ (dash-dot 
line) as a function of the scale factor $a$ after the end of inflation, when 
\mbox{$a=a_{\rm end}$} until today, when \mbox{$a=a_0$}. After the end of 
inflation the scalar field continues to dominate the Universe through its 
kinetic energy, because its energy density is initially much larger than the 
energy density of the background thermal bath, 
\mbox{$V_{\rm end}\gg \rho_B(a_{\rm end})\sim T_{\rm reh}^4$}.
However, during kination $\rho_\phi$ reduces faster than $\rho_B$ and, 
at \mbox{$a=a_*$}, the background density becomes dominant and the 
Hot Big Bang begins. Soon the kinetic energy of the scalar field is
reduced to zero and the field freezes when \mbox{$a=a_F$}, after
which, its energy density is kept constant and equal to $V_F$. 
The various types of attractor and frozen quintessence are shown depending on 
the position and steepness of the attractor solution (dashed line). 
In (a) the attractor meets the frozen field only after the present time, 
resulting in frozen quintessence and eternal acceleration. In (b) 
the case of a steep attractor is shown, with slope steeper than the background
density. Such an attractor meets the frozen field soon after the end of 
inflation and renders late $\phi$-domination of the Universe impossible. 
In (c) the case of the exponential attractor is shown, which mimics the
background density and results in a constant ratio between $\rho_\phi$ and
$\rho_B$. Eternal acceleration is avoided in this case. Finally, in (d) the
case of a mild attractor is shown, with slope not as steep as the background 
density. This attractor results in attractor quintessence with eternal 
acceleration.}
\label{q2}
\end{figure}

Thus, the above suggests
\begin{equation}
\phi(t)\simeq\left\{
\begin{array}{lrl}
\phi_F & \mbox{when}\hspace{0.5cm}
f(\phi_F)\gg F(t) & \;\mbox{\ Frozen}\\
 & & \\
\phi_{\rm atr}(t) & \mbox{when}\hspace{0.5cm}
f(\phi_F)\ll F(t) & \;\mbox{\ Attractor}
\end{array}
\right.
\end{equation}
where $\phi_{\rm atr}(t)$ is the solution of 
\begin{equation} 
f(\phi)\simeq F(t)\Leftrightarrow 
\frac{1}{4}(1+w_B)t^2 =-\int'\frac{d\phi}{V'(\phi)}
\label{atr}
\end{equation}
typically referred to as attractor solution~\cite{track}.
Because $F(t)$ is a growing function of time we expect that initially
the field remains frozen at $\phi_F$ until some time later when it
unfreezes and starts following the attractor solution (see Fig. \ref{q2}). 
This occurs when \mbox{$V(\phi_{\rm atr})\simeq V(\phi_F)$} so that, at all 
times,

\begin{equation}
V(\phi)=\mbox{min}\{V(\phi_F),V(\phi_{\rm atr})\}
\label{Vcons}
\end{equation}

Note that throughout the quintessential evolution of the scalar field
the field's acceleration $\ddot{\phi}$ is ignored because $\phi$, at most,
gently rolls down the ``quintessential tail'' of its potential. It should be 
pointed out here that this may not be quite true after unfreezing if the tail 
is steep.  However, as will be explained in \ref{sec:attractors-slope},
such case is not relevant for successful quintessence.

\subsubsection{Coincidence and acceleration requirements}
\label{sec:quint-requ}

Accounting for the fact that dark energy seems to begin dominating again 
today requires,

\begin{equation}
V(\phi_0)=\Omega_\phi\rho_0
\label{coin}
\end{equation}
where \mbox{$\phi_0\equiv\phi(t_0)$} is evaluated at the present time $t_0$ 
and \mbox{$\Omega_\phi\simeq 0.65$} is the ratio of $\rho_\phi(t_0)$ over the 
critical density today~\cite{Jaffe:2001tx}, 
\mbox{$\rho_0\simeq 10^{-120}m_P^4$}.

Moreover, as mentioned above, in order to explain the current accelerated 
expansion of the Universe we require, \mbox{$w_\phi<-\frac{1}{3}$}.
Finally, due to (\ref{Vcons}), the following constraint should be  satisfied,

\begin{equation}
\phi_0=\left\{
\begin{array}{llc}
\phi_F & & V(\phi_0)<V(\phi_{\rm atr}\mbox{\small $(t_0)$})\\
 & \Leftrightarrow & \\
\phi_{\rm atr}(t_0) & & 
V(\phi_F)>V(\phi_0)
\end{array}
\right.
\end{equation}

\section{How to model  quintessential inflation}
\label{sec:how-model-infl}

Apart from the requirements of inflation and quintessence discussed
above, quintessential inflationary models must satisfy some additional
generic conditions, such as
\begin{itemize}
\item {\bf ``Sterile'' inflaton:} It is necessary that the scalar
  field should not be coupled to the standard model fields so that it
  cannot decay into them at the end of inflation. This is so in order
  for the field to survive until today and play the role of
  quintessence.  This is most easily realized by having the scalar
  field be a gauge singlet.
\item {\bf Quintessential tail:} The scalar field should not have
  reached the minimum of its potential by the present time because it
  must have a tiny residual potential energy density today. Thus, the
  minimum of the potential should be placed rather far from the
  initial value (probably at infinity). Therefore, the required form
  of the potential contains two flat regions: the inflationary plateau
  and the quintessential tail.
\end{itemize}
Finally, it should be mentioned that, since quintessence is introduced
to dispense with the pure cosmological constant $\Lambda$, in quintessential
inflation we take \mbox{$\Lambda=0$} and \mbox{$V_{\rm min}=0$}.  A 
qualitative view of the form of the potential for Quintessential Inflation is 
shown in Fig. \ref{q1}. We now
give some further insights on quintessential and inflationary
evolution that will be useful in developing successful forms of the
scalar potential.

\subsection{Quintessence attractors and trackers}
\label{sec:quint-attr-track}

The quintessential tail requirement results in the existence of attractor 
solutions, which, when followed by the rolling scalar field, lead to dynamics 
and general behaviour independent of the initial conditions \cite{track}. 
These attractor solutions may be a benefit or a hazard as
regarding the quintessential requirements. Since the nature of the
attractor is decided by the specific form of the potential,
constraining the attractor is the first step towards the successful
modeling of quintessential inflation.

\subsubsection{Tracker solutions}
\label{sec:tracker-solutions}

A quintessential attractor solution is referred to as a ``tracker''
when it is such that it enables and assists attractor quintessence to
dominate the Universe at late times. This is achieved by having an
attractor which leads to late $\phi$-domination of $\rho_B$.  As a
consequence, when the scalar field begins the attractor phase of its
evolution, it will inevitably come to dominate the Universe and play
the role of quintessence.

The tracker concept is relevant only in attractor quintessence~(see
 Fig.~\ref{q2}(d)~), since in frozen quintessence
(Fig.~\ref{q2}(a)~) the attractor is never reached. As explained
above, after kination the field lies frozen with potential energy
$V_F$ much smaller than $\rho_B$ so that the Universe evolves according
to the SHBB. In attractor quintessence the attractor solution meets
the frozen field before the latter dominates the Universe so that, at
the onset of the attractor evolution \mbox{$V_{\rm atr}<\rho_B$}. Thus,
if the attractor is a tracker the ratio \mbox{$V_{\rm atr}/\rho_B$}
should be an increasing function of time (or, equivalently, of $\phi$).
From the attractor equation (\ref{atr}) and using also the equation of
motion of $\phi$ (\ref{infeqm}) we find,
\begin{equation}
\frac{m_P^2}{3(1+w_B)}\Big(\frac{V_{\rm atr}}{\rho_B}\Big)=
-V\int'\frac{d\phi}{V'}
\label{middle}
\end{equation}
Taking the time derivative of the above 
we find that the criterion for a tracker attractor is

\begin{equation}
\frac{d}{dt}\Big(\frac{V_{\rm atr}}{\rho_B}\Big)\geq 0\Leftrightarrow
V\geq (V')^2\int'\frac{-dV}{(V')^2}
\end{equation}
Similarly, it can be shown that

\begin{equation}
\Gamma\equiv\frac{VV''}{(V')^2}\geq 1\Leftrightarrow
\frac{d^2}{dt^2}\Big(\frac{V_{\rm atr}}{\rho_B}\Big)\geq 0
\end{equation}
which is usually enough for the kind of potentials used in
quintessence \cite{track}.  The case of equality in the above corresponds to
potentials with exponential quintessential
tail~\cite{exp-q}\cite{exp-new} and is the
limiting case of the tracker solution.  As we will discuss below,
tracker solutions exist only for potentials with tails of slope
milder than that of the exponential tail.
 
\subsubsection{Attractors and slope}
\label{sec:attractors-slope}

The behaviour of the attractor is strongly related to the slope of the
quintessential tail of the potential. This can be demonstrated as follows.
Derivating (\ref{middle}) with respect to $\phi$ and using also (\ref{eps}) we
obtain,
\begin{equation}
\frac{d(1/V_{\rm atr})}{d(1/\rho_B)}=\frac{2\varepsilon^2}{1+w_B}
\label{middle2}
\end{equation}

Let us parametrize the behaviour of \mbox{$\varepsilon(V)$} around the region
of $V$ of interest (the quintessential tail) as a power-law,

\begin{equation}
\varepsilon^2\equiv C\Big(\frac{1}{V}\Big)^q
\label{epsV}
\end{equation}
where $C$ is a positive constant.  The actual value of $\varepsilon^2(V\to 0)$
depends on the slope since, \mbox{$\varepsilon^2\propto (V')^2/V^2$} and
$V' \to 0$ as $V \to 0$. For the exponential slope
\mbox{$\varepsilon^2$=const.$>0$}.  For steeper slopes $(V')^2$ is larger than
the exponential case and thus, \mbox{$\varepsilon^2(V\!\to\!0)\to +\infty$}, 
whereas for milder slopes $(V')^2$ is smaller than the exponential case so that
\mbox{$\varepsilon^2(V\!\to\!0)\to 0$}.  Thus, in (\ref{epsV}) we may consider,

\begin{equation}
\begin{array}{rl}
q>0 & {\rm steep}\\
q=0 & {\rm exp}\\
q<0 & {\rm mild}
\end{array}
\end{equation}

Inserting (\ref{epsV}) into (\ref{middle2}) we find,

\begin{equation}
\frac{V_{\rm atr}}{\rho_B}=\frac{(1+w_B)V^q}{2C(1-q)}\propto 
\Big(\frac{1}{V}\Big)^{-q}
\end{equation}

From the above it is evident that for quintessential-tail with steep
\{mild\} slope $V_{\rm atr}$ reduces with time faster \{slower\} than
$\rho_B$. This means that, in the mild case the attractor is such that
$V_{\rm atr}$ approaches the background density and, therefore, it is
a tracker. In the steep case, however, if the attractor evolution
begins, there is no chance that the scalar field will come to dominate
the Universe and, therefore, such a case cannot result in
quintessential evolution. Thus, for potentials with steeper than
exponential tails only frozen quintessence is, in principle, possible.

One can imagine that, even if the onset of the steep attractor is
disastrous for quintessence, a steep-tail model may still be acceptable in
the case of frozen quintessence with the steep attractor beginning after
today. However, it turns out that a steep attractor begins very soon
after the end of inflation, which renders frozen quintessence in this
case also impossible to attain.  This can be understood as follows.
From (\ref{middle2}) after some trivial algebra  we get
\begin{equation}
\frac{1}{2}\frac{d}{d(\ln t)}\Big(\frac{\rho_B}{V_{\rm atr}}\Big)=
\frac{2\varepsilon^2}{1+w_B}-\frac{\rho_B}{V_{\rm atr}}
\label{middle3}
\end{equation}
where we also used \mbox{$\rho_B\propto t^{-2}$}. In the steep case, before 
the cross-over of $V_{\rm atr}$ and $\rho_B$, we have 
\mbox{$V_{\rm atr}\gg\rho_B$}. Also, in this case \mbox{$\varepsilon^2> 1$}.  
Thus, one can ignore the last term of (\ref{middle3}) and obtain,
\begin{equation}
\frac{\rho_B}{V_{\rm atr}}\simeq\int
\frac{4\varepsilon^2 d(\ln t)}{1+w_B} 
\end{equation}
The above suggests that, because $\varepsilon^2$ is very large, the time
required for the cross-over is small. Indeed, considering that in such
little time $\varepsilon^2$ does not change much, we may estimate the 
cross-over time as,
\begin{equation}
t_\times\simeq t_{\rm end}\exp\Big[\frac{1}{4}(1+w_B)\varepsilon^{-2}
\Big]\sim t_{\rm end}
\end{equation}
Thus, we see that the steep-tail attractor begins very soon after the
end of inflation and, therefore, one cannot achieve frozen
quintessence. The above are illustrated in Fig. \ref{q2}(b). 

It should be noted here that, for steep quintessential tails the $\ddot{\phi}$
term in the equation of motion cannot be ignored. The action of this term 
intensifies the rapid roll of the field and steepens the attractor even more.

\subsubsection{The exponential tail}
\label{sec:exponential-tail}

The exponential tail is the interface between steep and mild slopes.
In this case \mbox{$\varepsilon^2=$ const.} and (\ref{middle}) gives,
\begin{equation}
\frac{\rho_B(t)}{V_{\rm atr}(t)}=\frac{2\varepsilon^2}{1+w_B}={\rm constant}
\label{eps2}
\end{equation}
Therefore, we see that the scalar field attractor mimics the behaviour
of the background matter and evolves as \mbox{$V_{\rm atr}\propto t^{-2}$}.

The exponential tail is also marginally affected by the $\ddot{\phi}$
term in the equation of motion of the scalar field. If this term is
also taken into account the above equation becomes,

\begin{equation}
\frac{\rho_B(t)}{V_{\rm atr}(t)}=\frac{4\varepsilon^2}{1-w_B^2}={\rm constant}
\label{eps3}
\end{equation}
It is evident that the $\ddot{\phi}$ term is insignificant for milder
slopes, where slow--roll conditions are truly satisfied.

Since the attractor solution breaks down when \mbox{$\rho_B\sim V_F$} we see
that for frozen quintessence we must require,
\begin{equation}
\frac{V_{\rm atr}}{\rho_B}> 1\Leftrightarrow
\varepsilon^2<\frac{1}{4}(1-w_B^2)
\label{expfroz}
\end{equation}
If the above is violated we have attractor quintessence. In this case
we have \mbox{$V_{\rm atr}\propto\rho_B\propto t^{-2}$}. Also, using 
(\ref{H}) and the field equation (\ref{infeqm}) we find, 
\mbox{$\rho_{\rm kin}\equiv\frac{1}{2}\dot{\phi}^2= 
\varepsilon^2V_{\rm atr}^2/\rho_B\propto t^{-2}$}. Thus,
in overall \mbox{$\rho_\phi=\rho_{\rm kin}+V_{\rm atr}\propto t^{-2}$}, or
equivalently, \mbox{$\rho_\phi/\rho_B=$ constant} \cite{exp-q}. This, in view 
of (\ref{r-a}), implies that \mbox{$w_\phi=w_B\geq 0$}. Therefore, in 
attractor quintessence an exponential tail cannot lead to acceleration.
Moreover, in this case, because the attractor solution is such that
\mbox{$V_{\rm atr}<\rho_B$} at all times, the scalar field never comes to
dominate the Universe but remains a constant fraction of the overall
density (see Fig. \ref{q2}(c)~).  
Therefore, if the scalar field were to account for the dark
energy component of the late-time Universe then, at present, we should
demand that,

\begin{equation}
V_{\rm atr}\lsim\rho_B
\label{expatr}
\end{equation}
meaning that $V_{\rm atr}$ should be very close but not larger than $\rho_B$.

\begin{figure}
\begin{center}
\leavevmode
\hbox{%
\epsfxsize=4.0in
\epsffile{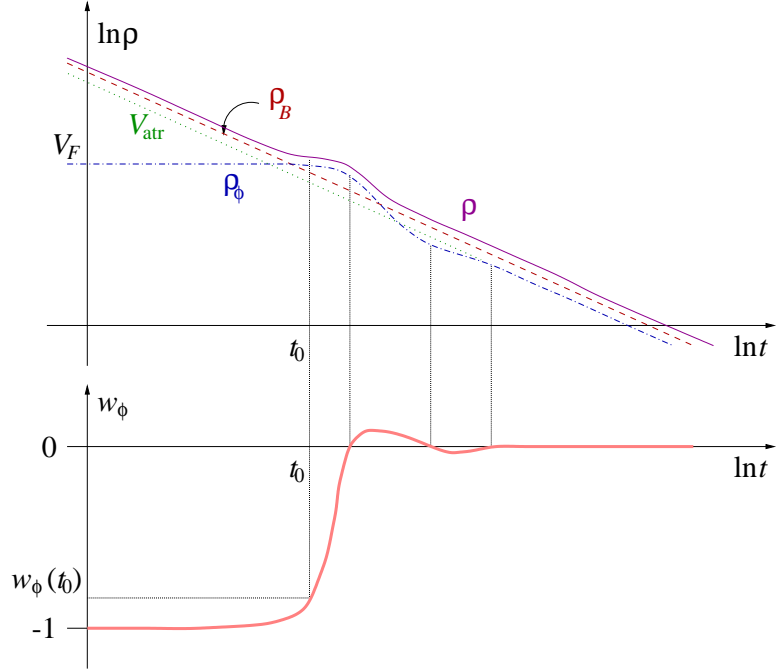}}
\end{center}

\caption{\footnotesize  Diagram illustrating the behaviour of the energy 
density $\rho_\phi$ of the scalar field as well as the equation of state 
parameter $w_\phi$ at the moment of unfreezing in the case of the exponential 
attractor, when coincidence is fixed near this moment. The upper graph shows 
the evolution of $\rho_\phi$ (dot-dash line) and of the background 
density $\rho_B$ (dashed line) near their cross-section point (it may be 
considered as a close-up of Fig. \ref{q2}(c)~). When $\rho_\phi$ meets the 
attractor solution $V_{\rm atr}$ (dotted line) it does not immediately 
unfreeze. Instead it remains ``superfrozen'' for a brief period, during which 
it dominates $\rho_B$ and results in a ``bump'' in the down-slope of the 
decreasing overall energy density $\rho$ (solid line). Then it 
oscillates briefly around the attractor before settling down to follow it. 
The above ``bump'' results in brief acceleration of the Universe expansion,
which is necessary to explain the current observations, while avoiding
eternal acceleration. During this transition a corresponding change of 
$w_\phi$ from -1 to 0 occurs, as shown in the lower graph. During the brief 
oscillation period of the system around the attractor, $w_\phi$ oscillates 
around zero. However, during the ``bump'' in $\rho$ the value of $w_\phi$ is 
still close to -1 because $\rho_\phi$, being ``superfrozen'', remains almost 
flat.}
\label{q4}
\end{figure}

It has been shown by numerical simulations that when the scalar field
unfreezes and begins to follow the attractor, its trajectory in phase
space, just after the transition, oscillates briefly around the
attractor evolution path (see Copeland {\em et al.} in \cite{exp-q}). 
This fact, in the case of the exponential
tail and when the attractor is really close to dominating the Universe
energy density (as suggested by (\ref{expatr})~), enables the field to
achieve some acceleration for a limited time interval until it settles
to the attractor (see Fig. \ref{q4}). Thus, attractor quintessence with 
exponential tail can result in brief but not eternal acceleration 
\cite{exp-new}.\footnote{In some cases of exponential quintessence
accelerating expansion may be terminated also due to gravitational 
backreaction \cite{Li:2001pk}.}

\subsubsection{The choice of quintessential tail}
\label{sec:choice-quint-tail}

String theory\footnote{actually, quantum gravity in general} 
considerations disfavor a Universe with eternal acceleration
because such a Universe features a future event horizon, which renders 
the $S$-matrix construction problematic because the asymptotic states of 
observables are not well defined \cite{strings}, similarly to the case of 
de-Sitter space \cite{Witten:2001kn}. Moreover, since the de-Sitter vacuum has 
finite temperature, a system cannot relax into a zero-energy supersymmetric 
vacuum while accelerating if the evolution is dominated by a single scalar 
field with a stable potential. If we take this constraint into consideration 
on our quintessence models then all of frozen quintessence, as well as 
mild-slope attractor schemes (Figs.~\ref{q2}(a) and \ref{q2}(d)~) become ruled
out. Thus, since steep-slope quintessential tails (Fig. \ref{q2}(b)~) have 
disastrous attractors, we conclude 
that only attractor quintessence with an exponential-tail is admissible for
quintessential evolution\footnote{Of course one can imagine quintessential 
tails which change slope from mild to steep, such as in 
\cite{hybridtail}. However, such features may be introduced 
only at the expense of invoking additional mass-scales and parameters, 
which we discard in our minimalistic approach to quintessential inflation.}.
Moreover, if we want to achieve brief acceleration we must consider
the unfreezing of the field near the present time. Therefore, the best
choice seems to be considering \mbox{$\phi_0\simeq\phi_F$} and also 
implementing the attractor constraint (\ref{expatr}).

Thus, in search of a model that communicates information from the
inflationary era down to the quintessential tail (i.e. a model without
phase transitions or a multi-branch potential), we are led to try a
quintessential tail of the form,
\begin{equation}
V(\phi\gg\phi_{\rm end})\simeq V_{\rm end}\exp(-\lambda\phi/m_P)
\label{exp}
\end{equation}
where \mbox{$\lambda>0$} is a parameter. Using (\ref{eps}) we obtain
\mbox{$\varepsilon=-\lambda/\sqrt{6}$}.  Hence, in view of (\ref{expatr}) we 
find for $\lambda$,
\begin{equation}
\lambda\gsim\sqrt{3/2}
\label{l}
\end{equation}
meaning that $\lambda$ should be close but not smaller than $\sqrt{3/2}$.
Using this we may employ the coincidence requirement (\ref{coin}), using 
(\ref{ffgr}) for \mbox{$\phi_0=\phi_F$}, to find the corresponding value 
for $V_{\rm end}$. However, if one does that then it can be shown that, 
in this case, it is impossible to satisfy the BBN constraint (\ref{bbn}). 
This is because the slope of the exponential tail when $\lambda$ is given 
by (\ref{l}) is not sufficient to achieve coincidence and
simultaneously retain a high value of $V_{\rm end}$, which would have
enabled kination to terminate before the onset of BBN.

Fortunately, there is a way out from this disappointing result. What
one may do is attempt to modify the quintessential tail in such a way
that the attractor may be preserved, more or less intact but the
magnitude of the potential energy density of the scalar field become
nevertheless additionally suppressed, compared to the pure exponential
tail. This is possible to achieve if one combines the exponential tail
with another tail of milder slope. The obvious candidate is the
Inverse Power-Law (IPL) tail, which is rather popular in quintessence
models and may even have some theoretical
motivation~\cite{susy-q}.
Thus, the suggested quintessential tail is of the form,

\begin{equation}
V(\phi\gg\phi_{\rm end})\simeq 
V_{\rm end}\frac{\exp(-\lambda\phi/m_P)}{(\phi/m)^k}
\label{tail}
\end{equation}
where \mbox{$k\geq 1$} is an integer and \mbox{$m<m_P$} is a mass-scale 
characteristic of the IPL-slope. The above quasi-exponential quintessential 
tail has the same attractor solution with the pure exponential one for 
large values of $\phi$. Indeed, from (\ref{ffgr}) we see that 
\mbox{$\phi_F\gg m_P$}, which suggests,

\begin{equation}
V'(\phi_F)=-\frac{V}{m_P}\Big[\lambda+k\Big(\frac{m_P}{\phi_F}\Big)\Big]
\simeq-\frac{\lambda V}{m_P}
\end{equation}
i.e. $V'$ approaches the value of the pure exponential tail (corresponding 
to \mbox{$k=0$}). Thus, since the form of the attractor solution (\ref{atr}) 
is decided by $V'$ and not by $V$, we find that the exponential-tail attractor 
is attained while there is an additional suppression of the potential
energy due to the contribution of the IPL factor.\footnote{Note here that 
there is no danger to reach the IPL attractor, which is valid when
\mbox{$\phi\ll m_P$}, because this attractor is of milder slope and it is 
therefore impossible to reach when coincidence occurs for 
\mbox{$\phi_0=\phi_F\gg m_P$}.}

It is evident that, given a large enough $V_{\rm end}$ (to satisfy the
BBN constraint), coincidence is attained with low values of $k$ only
when using adequately low values for $m$.

\subsection{Insights for the inflationary era}
\label{sec:insights-infl-era}

It is easy to see that the exponential behaviour of the potential at
the quintessential tail cannot carry over to the inflationary era
because of the steepness it results to. Indeed, suppose that the IPL
contribution is somehow eliminated at the inflationary part of the
potential so that during inflation $V(\phi)$ is given by (\ref{exp}).
Then in view of (\ref{l}) and using (\ref{eps}) we find , \mbox{$\eta\gsim 1$} 
i.e. the second slow--roll condition cannot be satisfied and
it is impossible to achieve inflation. Obviously any IPL contribution
would only make matters worse. Thus, both the exponential and the IPL
features have to be modified during the inflationary era. This
modification is not trivial.

The prime obstacle is the huge difference between the energy scales of
the inflationary plateau and the quintessential tail, 
\mbox{$V_{\rm end}\gg V_F\sim\rho_0$} as required by BBN and coincidence. 
Given that BBN requires, roughly, that \mbox{$V_{\rm end}\gsim 10^{13}$GeV}, 
a steep inflationary plateau results either in very brief inflation,
which cannot account for the horizon and flatness problems, or, if not
taken to be brief, in strongly super-Planckian inflationary energy
scale. A flat inflationary plateau, however, because it has to
``prepare'' for the deep dive towards $V_F$ after the end of
inflation, typically features a large value of $|V''|$, which, through
$\eta$, results in unacceptably large values of the spectral index $n$
[c.f.~(\ref{n})]. Thus, it is not at all easy to design
a successful potential for quintessential inflation, let alone using
few mass scales and parameters. Nevertheless, it is not impossible as
we show in the following example.

\section{A Concrete Example}
\label{sec:examples}

Following the insights given in Sec. \ref{sec:how-model-infl} we now
present a concrete model realization in terms of a specific form of
the scalar potential and demonstrate how it meets all the necessary
requirements which have been outlined in Sec.~\ref{sec:infl-quint}.

\subsection{The model}

\begin{figure}
\begin{center}
\leavevmode
\hbox{%
\epsfxsize=5.0in
\epsffile{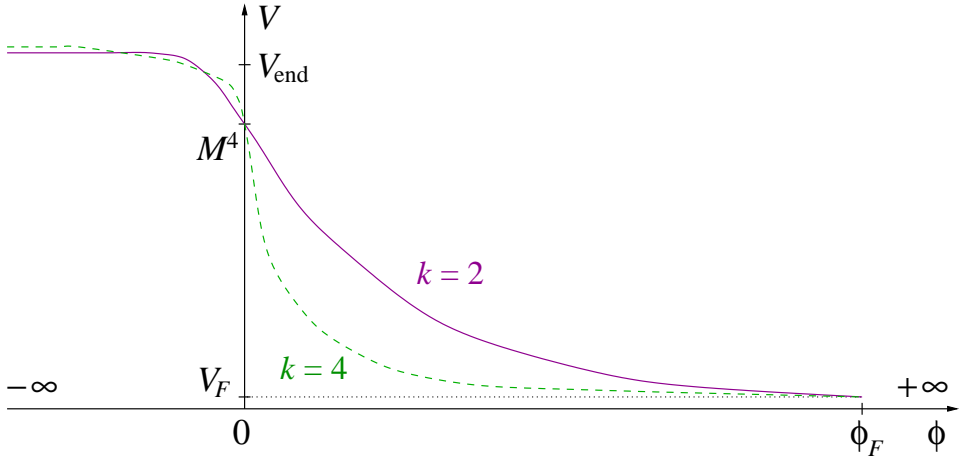}}
\end{center}

\caption{\footnotesize Artist's view of the potential (\ref{model}) for 
\mbox{$k=2,4$} (solid line and dashed lines respectively). 
The decrease between the inflationary plateau and the 
quintessential tail corresponds to a drastic reduction of the energy density 
by more than a hundred orders of magnitude. However, the differences on
the inflationary energy scale, which depend on $k$ are only a few factors of 
2. All the models of the family cross at the value \mbox{$V(\phi=0)=M^4$} and 
\mbox{$V_F=V(\phi_F)$}.} 
\label{q3}
\end{figure}

Consider the potential,
\begin{equation}
V(\phi)=M^4[1-\tanh(\phi/m_P)]
\Big[1-\sin\Big(\frac{\pi\phi/2}{\sqrt{\phi^2+m^2}}\Big)\Big]^k
\label{model}
\end{equation}
where the scalar field $\phi$ lies in the range \mbox{$-\infty<\phi<+\infty$},
\mbox{$M,\:m<m_P$} are two mass parameters and \mbox{$k>0$} is an
integer. This is schematically illustrated in Fig.~\ref{q3}. The
above may be written in terms of the dimensionless field variable

\begin{equation}
x\equiv\frac{\phi}{m}
\label{x}
\end{equation}
as follows,
\begin{equation}
V(x)=M^4[1-\tanh(zx)]
\Big[1-\sin\Big(\frac{\pi x/2}{\sqrt{1+x^2}}\Big)\Big]^k
\label{Vx}
\end{equation}
where 
\begin{equation}
z\equiv\frac{m}{m_P}
\label{z}
\end{equation}
Typically, \mbox{$0<z\ll 1$}.  For negative values of $x$ the above
becomes,
\begin{equation}
V(x\ll 0)\simeq 2^{k+1}M^4(1-e^{2zx}-\frac{k\pi^2}{64x^4})\simeq 
2^{k+1}M^4
\label{Vinf}
\end{equation}
i.e. it approaches a constant, non-zero, false vacuum energy density
which is responsible for driving the inflationary era.  On the other
hand, for very large positive values of $x$ the potential (\ref{Vx})
becomes,
\begin{equation}
V(x\gg 0)\simeq 2^{1-k}(\frac{\pi}{4})^{2k}e^{-2zx}\frac{M^4}{x^{4k}}
\label{Vq}
\end{equation}
which has the desired form of the quasi-exponential tail 
appropriate for quintessence. Note that, in view of
(\ref{z}) and comparing the above with (\ref{tail}) we see
that \mbox{$\lambda = 2$}, a value which is in good agreement with (\ref{l}).  
From (\ref{Vx}) we find,
\begin{equation}
V'=-\frac{V}{m}\,W \hspace{0.4cm}\mbox{where}\hspace{0.4cm}
W(x)\equiv \frac{2z}{1+e^{-2zx}}+\frac{k\pi}{2}\tan
\Big[\frac{\pi}{4}\Big(1+\frac{x}{\sqrt{1+x^2}}\Big)\Big](1+x^2)^{-3/2}
\label{V'}
\end{equation}
and also,
\begin{eqnarray}
V'' & = & \frac{V}{m^2}\,\left(W^2-\frac{dW}{dx}\right) 
\hspace{2cm}\mbox{where}\label{V''}\\
 & & \nonumber\\
\frac{dW}{dx} & = & \frac{z^2}{[\cosh(zx)]^2}+
\frac{2k(\pi/4)^2}{(1+x^2)^3}
\left\{
\cos\Big[\frac{\pi}{4}\Big(1+\frac{x}{\sqrt{1+x^2}}\Big)\Big]
\right\}^{-2}\hspace{-0.4cm}
-\frac{3k\pi x}{2(1+x^2)^{5/2}}
\tan\Big[\frac{\pi}{4}\Big(1+\frac{x}{\sqrt{1+x^2}}\Big)\Big]\nonumber
\end{eqnarray}

\subsection{Inflation}

Inflation corresponds to \mbox{$\phi\ll 0$} or, equivalently, \mbox{$x\ll
  0$}. From (\ref{eps}), (\ref{V'}) and (\ref{V''}) we
find that, in this region 
 \begin{eqnarray}
\varepsilon(x\ll 0) & \simeq & -\frac{1}{\sqrt{6}}\Big(\frac{m_P}{m}\Big)
\Big(2ze^{2zx}-\frac{k\pi^2}{16x^5}\Big)\nonumber\\
 & & \label{neweps}\\
\eta(x\ll 0) & \simeq & -\frac{1}{3}\Big(\frac{m_P}{m}\Big)^2
\Big(4z^2e^{2zx}+\frac{5k\pi^2}{16x^6}\Big)\nonumber
\end{eqnarray}

It can be checked that, in the above, the exponential term is dominant
in the inflationary region. In view of this fact one finds that
inflation is terminated when the second slow-roll condition is
violated, i.e.  \mbox{$|\eta(x_{\rm end})|=1$}, where 
\mbox{$x_{\rm end}\equiv\phi_{\rm end}/m$}. Hence, we find,

\begin{equation}
\phi_{\rm end}\simeq-\ln(2/\sqrt{3})\,m_P
\label{xend}
\end{equation}
Thus, we see that $\phi_{\rm end}$ is independent of the mass scales and of 
$k$.

Inserting the above into (\ref{Vinf}) we obtain,

\begin{equation}
V_{\rm end}\simeq 2^{k-1}M^4
\label{Vend}
\end{equation}

Now, using (\ref{N}), we find,

\begin{equation}
\phi_N\equiv\phi(N)=-\frac{1}{2}\ln[4(N+1/3)]\,m_P
\label{xN}
\end{equation}
which is again independent of $M,m$ and $k$.

Using the above (\ref{dr/r}) and (\ref{cobe}) provide the COBE constraint,

\begin{equation}
\frac{2^{\frac{k+1}{2}}}{\sqrt{3}\,\pi}\Big(\frac{M}{m_P}\Big)^2
(N_{\rm dec}+1/3)=10^{-5}
\label{COBE}
\end{equation}

This constraint determines the magnitude of the inflationary energy scale $M$ 
in terms of $k$ and of the reheating efficiency factor $\alpha$. Using 
(\ref{NTH}) and taking \mbox{$\alpha\simeq 1/100$} and 
\mbox{$k\leq 4$} we find,

\begin{equation}
M\sim 10^{15} \: \mbox{GeV}
\label{M}
\end{equation}
which is roughly the scale of grand unification
similarly to the cases of Chaotic, Hybrid and Natural inflation. 

Let us also calculate the spectral index of the density perturbation spectrum.
Using (\ref{n}), (\ref{neweps}) and (\ref{xN}) we find,

\begin{equation}
n(N)-1\simeq-\frac{6}{3N+1}\Big[1+\frac{9}{8(3N+1)}\Big]
\end{equation}
which is independent of $k$ and of the inflationary energy scale $M$. With
\mbox{$\alpha\simeq 1/100$} and using (\ref{Ns}) the above gives,

\begin{equation}
n(N_{\rm dec})\approx n(N_{\rm eq})\simeq 0.97
\end{equation}
which is in excellent agreement with observations \cite{cmbobs}.

Using (\ref{Hinf}), (\ref{Trehg}), (\ref{Vend}) and 
(\ref{COBE}) one finds,

\begin{equation}
T_{\rm reh}\simeq\frac{\alpha}{4}\frac{10^{-5}m_P}{\sqrt{N_{\rm dec}+1/3}}
\sim 10^{9} \: \mbox{GeV}
\label{Treh}
\end{equation}
which actually saturates the gravitino bound. Finally, from (\ref{T*}),
(\ref{Vend}) and (\ref{COBE}), while also using \mbox{$g_*=106.75$}, one 
obtains,

\begin{equation}
T_*\simeq\frac{\alpha^3\pi^2}{96}\sqrt{\frac{g_*}{30}}
\frac{10^{-10}m_P}{(N_{\rm dec}+1/3)^2}\sim 10\;\mbox{MeV}> T_{\rm\sc bbn}
\label{T**}
\end{equation}
i.e. the BBN constraint is satisfied. Note that the above estimates
of $T_{\rm reh}$ and $T_*$ are $k$-independent.

Finally, evaluating (\ref{NF}) and (\ref{Ns}) one finds that
\mbox{$N_F\simeq 57$} and \mbox{$N_H\simeq 74$} respectively, which, using 
(\ref{xN}), suggests that the horizon and flatness problems are solved if 
the initial value of the field is,

\begin{equation}
\phi_{\rm in}<-\frac{1}{2}\ln[4(N_H+1/3)]\,m_P\simeq-2.85\,m_P
\label{initial}
\end{equation}
which is natural to satisfy if one considers that we expect
\mbox{$|\phi_{\rm in}|\sim M_P$}. Thus, no fine-tuning of initial conditions 
for inflation is required.

\subsection{Quintessence}

The quintessential region of the potential (\ref{Vx}) corresponds to
\mbox{$\phi\gg 0$}. According to (\ref{Vq}) the quintessential tail is,

\begin{equation}
V(\phi\gg 0)\simeq 2^{1-k}(\frac{\pi}{4})^{2k}e^{-2\phi/m_P}M^4
\Big(\frac{m}{\phi}\Big)^{4k}
\label{Vquint}
\end{equation}

Let us find the attractor first. Using (\ref{f}) we find,

\begin{equation}
f(\omega)=\frac{2^{k-3}m^2}{\pi^{2k}z^{2(2k+1)}M^4}\times I(\omega)
\hspace{1cm}\mbox{with}\hspace{1cm}\omega\equiv 2\phi/m_P
\end{equation}
where 

\begin{equation}
I(\omega)\equiv\int'\frac{e^\omega\omega^{4k}d\omega}{1+\frac{4k}{\omega}}
\simeq e^\omega\omega^{4k}\hspace{1cm}\mbox{when}\hspace{1cm}\omega>4k
\label{I}
\end{equation}

In this case we may use (\ref{wD}) and  (\ref{eps3}) to obtain the attractor solution,

\begin{equation}
V_{\rm atr}(t)=\frac{1}{2}\Big(\frac{1-w_B}{1+w_B}\Big)
\Big(\frac{m_P}{t}\Big)^2\propto t^{-2}
\label{attr}
\end{equation}

Thus, in view of (\ref{wD}) and (\ref{z}) we obtain,

\begin{equation}
\frac{V_{\rm atr}}{\rho_B}=\frac{3}{8}\,(1-w_B^2)
 \hspace{1cm}\mbox{for}\hspace{1cm}\phi_{\rm atr}(t)>2k\,m_P
\end{equation}
which is in agreement with (\ref{expatr}).

As explained in Sec. \ref{sec:choice-quint-tail}, in order to achieve a 
brief period of acceleration (to account for the SN Ia observations) but not 
eternal acceleration (disfavored by string theory considerations) we 
need to have attractor quintessence beginning today, i.e. 
\mbox{$\phi_{\rm atr}(t_0)=\phi_F$}. In this case, since 
\mbox{$w_B(t_0)=w_m=0$}, we have,

\begin{equation}
V_F=V_{\rm atr}(t_0)=\frac{3}{8}\rho_B(t_0)=\frac{3}{11}\,\rho_0
\label{coin2}
\end{equation}
where \mbox{$\rho_0=\rho_B(t_0)+V_F$}. Hence, we see that, because
\mbox{$\Omega_\phi\approx 2/3$} coincidence, as defined in (\ref{coin}), is 
not quite attained when the attractor is met. Indeed, for coincidence we need
\mbox{$V_F=2\rho_B(t_0)$}.  However, if this were so, because
\mbox{$V_{\rm atr}(t_0)=V_F$} we would have had eternal acceleration,
according to (\ref{expfroz}). Still, coincidence may be attained
briefly because of the oscillations of the scalar field evolutionary
track around the attractor (\ref{attr}), just after the unfreezing of
the field. These oscillations are due to the fact that the field
retains its frozen value for a little while after crossing over the
attractor's path, being effectively ``superfrozen''. This may increase
somewhat the ratio $V_F/\rho_0$ before the field actually unfreezes.
Since we only require an increase of the order of 
\mbox{$\sim\frac{\Delta \Omega_\phi}{\Omega_\phi}\approx 3/5$}, 
we feel that coincidence is realistically possible to achieve. 
Moreover, while the field is ``superfrozen'' we
expect that \mbox{$w_\phi\approx-1$}. However, in order to obtain a specific 
value of $w_\phi$ numerical simulations are necessary. Still, we are confident
that successful coincidence can be attained by minute adjustments of the 
mass scales, which suggests that our treatment is sufficient for order of
magnitude estimates of $m,M$ (see also Sec.~\ref{sec:disc-concl}).

In the following we will use (\ref{coin2}) to estimate the mass-scale $m$.
The frozen value of the field is
estimated by (\ref{ffgr}), (\ref{xend}) and (\ref{Vend}) to be,

\begin{equation}
\phi_F=\sqrt{6}\left\{\frac{2}{3}-\frac{1}{\sqrt{6}}\ln(2/\sqrt{3})+
\ln\left[\frac{24}{\alpha^2}\sqrt{\frac{10}{g_*}}
(N_{\rm dec}+1/3)10^5\right]\right\}
m_P
\label{zx}
\end{equation}
where we have also used (\ref{COBE}).  Note that $\phi_F$ is independent
of the mass scales $M,m$ and of $k$, similarly to $\phi_{\rm end}$.  For
\mbox{$\alpha\simeq 1/100$} and \mbox{$g_*=106.75$} the above suggests that
\mbox{$\phi_F\simeq 67\,m_P$}, i.e. \mbox{$\phi_F/m_P> 4k$} \ for 
\mbox{$k\leq 4$} and we are, therefore, safely into the region of the
exponential attractor, as previously assumed in (\ref{I}).

Inserting (\ref{zx}) into (\ref{Vq}) and employing (\ref{coin2}), after a 
little algebra we find,

\begin{eqnarray}
m & = & 4\sqrt{3\pi}\left\{
\frac{2}{3}-\frac{1}{\sqrt{6}}\ln(2/\sqrt{3})+\ln\left[
\frac{24}{\alpha^2}\sqrt{\frac{10}{g_*}}(N_{\rm dec}+1/3)10^5\right]\right\}
\;\times\nonumber\\
 & & \\
 & \times & \left\{\frac{3}{44\pi^2}(N_{\rm dec}+1/3)^2 10^{10}
\left[
\frac{24e^{2/3}}{\alpha^2}\sqrt{\frac{10}{g_*}}(N_{\rm dec}+1/3)10^5
\right]^{2\sqrt{6}}\right\}^{1/4k}\Big(\frac{\rho_0^{1/4}}{m_P}\Big)^{1/k}
m_P\nonumber
\end{eqnarray}
where we used (\ref{z}) and (\ref{COBE}). With 
\mbox{$\alpha\simeq 1/100$} and \mbox{$g_*=106.75$} the above becomes,

\begin{equation}
m\sim 4\times 10^2(3\times 10^{70})^{1/4k}10^{-30/k}\,m_P
\end{equation}

Thus, we see that the highest the value of $k$ the largest the 
magnitude of $m$. This is easily understood 
as follows. Since $k$ determines the IPL decrease of the potential energy
$V$ of the scalar field in the quintessential part of its evolution, a
large $k$ results in substantial suppression of $V$. This allows $m$,
to which $V$ is proportional [c.f. (\ref{Vquint})], to be large. In the
opposite case, however, the decrease of $V$ is not as substantial. As
a result $m$ has to be small so that $V$ may manage to drop as low as
$\rho_0$.

In the next table we show the values of $m$ for small values of $k$.

\bigskip

\begin{center}
\begin{tabular}{|c||c|c|c|c|}
\hline
$k$ & 1 & 2 & 3 & 4 \\ \hline
$m$ & $10^8$~GeV & $10^{15}$~GeV & $10^{17}$~GeV & $10^{18}$~GeV\\
\hline
\end{tabular}
\end{center}

\bigskip

From this table it is clear that we should consider only \mbox{$k\leq 4$}
in order to avoid super-Planckian values of $m$, which would result by
higher values of $k$.  Note also that for \mbox{$m\sim m_P$} we have
\mbox{$\phi_{\rm end},\phi_F={\cal O}(1)\times m_P$} and, therefore, the
approximations used to obtain (\ref{Vinf}) and (\ref{Vq}) are
challenged.  Still, we believe that even in this case our treatment
remains valid for order-of-magnitude estimates, especially for
quantities like $T_*, T_{\rm reh}, n$ or even $M$, which are
essentially independent of $k$. From the above
table we notice also that for \mbox{$k=2$} we may identify
\mbox{$m=M$}. Similarly, for \mbox{$k=4$} we may consider
\mbox{$m=m_P$}.  Thus, in these cases {\em we attain successful 
quintessential inflation with the use of only two, natural mass 
scales, the Planck mass and the scale of grand unification}. These two
models are:

\begin{itemize}
\item
{\bf Model 1:}

\begin{equation}
V(\phi)=M^4[1-\tanh(\phi/m_P)]
\Big[1-\sin\Big(\frac{\pi\phi/2}{\sqrt{\phi^2+M^2}}\Big)\Big]^2
\end{equation}

\item
{\bf Model 2:}

\begin{eqnarray}
V(\phi) & = & M^4(\phi)\;[1-\tanh(\phi/m_P)]\nonumber\\
 & & \\
M(\phi) & \equiv & 
M\Big[1-\sin\Big(\frac{\pi\phi/2}{\sqrt{\phi^2+m_P^2}}\Big)\Big]
\nonumber
\end{eqnarray}

\end{itemize}
Both these models provide inflation at the scale of grand unification,
with reheating temperature \mbox{$T_{\rm reh}\sim10^{9}$GeV} and spectral
index of the density perturbation spectrum \mbox{$n\simeq 0.97$}. Also,
since the field unfreezes at the present time we expect a brief period
of acceleration with \mbox{$w_\phi\approx -1$}.

However, to obtain a precise value of 
$w_\phi$ one would have to employ numerical simulations. In the first of 
\cite{exp-new} such numerical simulations suggest that successful coincidence
may be achieved with \mbox{$\lambda\approx 3/2$}. We can obtain this in our 
model if we take \mbox{$m_P\rightarrow \frac{4}{3}\,m_P$} in (\ref{model}). 
The subsequent results are largely unaffected, except the value of $m$ which 
is reduced, as expected. It turns out that, in this case,
one can have \mbox{$m\sim M\sim 10^{15}$GeV} for \mbox{$k=4$}. Thus, according 
to those numerical simulations, another
successful quintessential inflationary model would be,

\bigskip

$\bullet$ {\bf Model 3:}

\nopagebreak

\begin{equation}
\begin{array}{lll}
V(\phi) & = & M^4(\phi)\;
[1-\tanh(\frac{3}{4}\phi/m_P)]\\
 & & \\
M(\phi) & \equiv & 
M\Big[1-\sin\Big(\mbox{\large
$\frac{\pi\phi/2}{\sqrt{\phi^2+M^2}}$}\Big)\Big]
\end{array}
\end{equation}

\bigskip

Note that Model~1 of the above models may be constructed in the context of
the theoretical framework of \cite{Kaganovich:2001fc} by considering the 
pre-potentials, \mbox{$V_1(\phi)\equiv M^4\sin\Big($}\mbox{\large 
$\frac{\pi\phi/2}{\sqrt{\phi^2+M^2}}$}$\Big)$ and
\mbox{$V_2(\phi)\equiv \frac{1}{4}M^4/[1-\tanh(\phi/m_P)]=
\frac{1}{8}M^4(e^{2\phi/m_P}+1)$}.

\section{Discussion and conclusions}
\label{sec:disc-concl}

We have developed general criteria required in unified frameworks for
inflation and quintessence in which a single scalar field is chosen to
drive both. By using such a minimal field content we avoid having to 
fine-tune couplings and mass-scales.  Moreover, the initial conditions 
for quintessence are fixed at the end of the inflationary epoch.

Quintessential inflation is realized if the scalar potential features two
flat regions, the inflationary plateau and the quintessential tail. Inflation
in such models is of the non-oscillatory type, in which reheating is achieved
due to the gravitational production of particles at the end of the 
inflationary period. Such reheating is inefficient and leads to a brief 
period of kination, which has to be terminated before Big Bang 
Nucleosynthesis. After kination the Universe evolves according to the 
Standard Hot Big Bang, while the scalar field lies frozen until the present 
epoch, when, once more, its potential energy density becomes 
comparable with the background density of the Universe, leading to the 
currently observed accelerated expansion.

Our treatment suggests that, in order to achieve a brief acceleration period 
occurring at present the quintessential tail of the scalar potential should
be of quasi-exponential form. Also, the inflationary plateau should be 
such that the deep drop towards the quintessential tail does not result in 
an excessive value for the spectral index of density perturbations and CMBR 
anisotropies. We have demonstrated that the above are, indeed, possible to 
achieve and have presented and analyzed a class of successful models.  
In our quintessential inflationary models the inflationary energy scale is 
of the order of the scale of grand unification, similarly to Hybrid, 
Chaotic and Natural inflation. The reheating temperature, however, is low 
enough not to violate the gravitino constraint. Moreover, inflation manages 
to satisfy the COBE normalization constraint, to easily account for the
horizon and flatness problems and, finally, to result in a CMBR spectral 
index \mbox{$n \approx 0.97$}, in remarkable agreement with the 
latest CMBR data. As far as the quintessential part is concerned our models
lead to a brief acceleration period by unfreezing the scalar field 
at present. Because the field is briefly ``superfrozen'' we expect that 
\mbox{$w_\phi\approx -1$} as required. 

Two of our models are singled out because they achieve the desired results
with the use of only two, natural mass scales, namely the Planck mass and the
grand unification energy scale. Thus, these models do not introduce any
additional fine tunning for quintessence and in that respect they outshine the
cosmological constant alternative. 

Similar to other models of quintessence, ours also requires the expectation 
value of the field today to be of order the Planck mass. This fact is an 
indicator that we may realize our model in the context of supergravity. 
Indeed, hidden sector fields in supergravity and/or superstring models offer, 
potentially natural, candidates for our ``double-task'' scalar field $\phi$. 
Other alternatives include moduli fields or the radion field in models with 
large extra dimensions. Seeking for a detailed derivation of the possible 
forms of the scalar potential from a ``Theory of Everything'' lies however 
outside the scope of our present paper, which is mainly phenomenological.
Nevertheless, we find it remarkable that such quintessential
inflationary potentials, meeting all the required constraints and
requirements, are possible to construct with the use of only one
scalar field and very few natural mass scales.

\bigskip

\noindent
{\Large\bf Acknowledgments}

\bigskip

Work supported by Spanish grant PB98-0693 and by the European
Commission RTN network HPRN-CT-2000-00148

\end{document}